\documentclass[runningheads]{llncs}
\usepackage{array}
\usepackage[T1]{fontenc}
\usepackage{graphicx}
\usepackage{booktabs}
\usepackage[misc]{ifsym}
\usepackage{url}

\newcommand{\corr}{(\Letter)}

\usepackage{mwe}
\usepackage{booktabs}

\makeatletter
\let\orig@maketitle\maketitle
\renewcommand{\maketitle}{%
  \orig@maketitle
  \thispagestyle{firstpageaccepted}%
}
\newcommand{\ps@firstpageaccepted}{%
  \renewcommand{\@oddhead}{\small\itshape\hfil Accepted at HLDM@ECMLPKDD Conference, Naples, Italy, 2026 (pre-print)\hfil}%
  \renewcommand{\@evenhead}{}%
}
\makeatother

\begin{document}

\title{Understanding Operator Attitudes Toward AI-Supported Decision Making in Maritime Operations}

\titlerunning{Understanding Operator Attitudes}


\author{Doreen Jirak\inst{1} \corr \orcidID{0009-0003-2839-3475}  \and Armeen Saroukanoff \inst{2} \and Dirk van Rooy \inst{1}  \orcidID{0000-0003-2525-5408}}

\authorrunning{D. Jirak et al.}

\institute{University of Antwerp, Paardenmarkt 94, 2000 Antwerp, Belgium \email{\{doreen.jirak,dirk.vanrooy\}@uantwerpen.be}
\and
Antwerp Maritime Academy (AMA), Noordkasteel Oost 6, 2030 Antwerp, Belgium \email{armeen.saroukanoff@hzs.be}
}

\maketitle   

\begin{abstract} 
Maritime Autonomous Surface Ships (MASS) and AI-\\supported decision assistants are expected to transform maritime operations, but their safe integration depends on how maritime professionals perceive and trust such systems. This paper presents a survey study on maritime stakeholders’ attitudes toward an AI-supported assistant in collision-avoidance scenarios. Participants evaluated technology anxiety, trust in automation, and explanation quality using established and adapted questionnaires, complemented by sentiment and thematic analysis of open-ended responses Results indicate a generally positive disposition toward maritime technology, no clear age-related differences in openness, stable trust across scenarios, and more scenario-sensitive, multidimensional explanation ratings. Open responses showed that participants valued support for decision-making, situation awareness, and confidence-building, while raising concerns about AI reliability, over-reliance and loss of expertise. The findings suggest that maritime AI systems should not focus solely on increasing automation or trust, but on supporting calibrated reliance through transparent, reliable, and operationally meaningful design with domain experts in the loop.

\end{abstract}
\section{Introduction}
Maritime Autonomous Surface Ships (MASS) are increasingly discussed as a transforming development in maritime transport, with the potential to improve safety, and operational efficiency and resilience. Advances in sensing, artificial intelligence (AI), and decision-support systems enable vessel agents to perceive the environment, plan routes, perform collision avoidance \cite{Zhang21}, and support navigational decision-making, aligning with humans' situational awareness (SAW) \cite{Endsl23}. Similar to autonomous driving, the IMO\footnote{\url{https://www.imo.org/en/mediacentre/hottopics/pages/autonomous-shipping.aspx}} defined L1-L4 autonomy levels, where L1 represents the current state at sea with decision support and humans on board while the L4 stage describes a fully remote vessel with no human intervention. Implementing these different levels creates a socio-technical transition in which human operators, autonomous systems, and regulatory frameworks must co-evolve.

A central challenge is that increasing autonomy does not remove humans from maritime operations \cite{Li24}, but changes their roles \cite{DeVos21,Merwe26}. Officers of the watch (OOW), Captains, and remote operators may remain responsible for monitoring system behavior, interpreting recommendations, supervising autonomous functions, and intervening when necessary \cite{Veitc22}. This creates a tension between automation’s promise to reduce workload and mitigate human error, and the continued expectation that operators understand, trust, supervise, and override autonomous systems when required. Moreover, attributing accidents to human error alone risks oversimplifying the problem \cite{Sanch21}. Autonomous decision support should therefore be understood not as replacing humans with objective machines, but as redistributing tasks, authority, and responsibility between humans and automation \cite{Murra22}.

For MASS and AI-supported navigation \cite{Lyu24,Veitc24,Durli25}, this redistribution raises key questions for human-machine decision-making. Operators may value systems that integrate information and reduce cognitive burden, while remaining reluctant to delegate final authority in safety-critical situations involving collision avoidance, COLREG interpretation (Convention on the International Regulations for Preventing Collisions at Sea), and uncertain sensor data. Trust is therefore central, but safe collaboration requires calibrated rather than unconditional trust, i.e. operators should rely on automation when appropriate, while remaining able and willing to question or override it when necessary.

Explainable AI (XAI) is often proposed as a mechanism to support such calibrated trust \cite{Merwe23,Merwe24}. In maritime operations, explanations may help operators understand why a maneuver is recommended, how traffic situations are interpreted, whether navigational rules are followed, and which uncertainties or alternatives were considered \cite{Dhano21}. However, explanations may also increase cognitive load if they are poorly timed, overly technical, or misaligned with the operator’s task context \cite{Malla23}. This is especially critical on the vessel bridge, where situation awareness must be maintained under time pressure and decision support must fit established operational routines.
Against this background, it is necessary to better understand how maritime stakeholders perceive autonomous decision-support systems \cite{Yoshi25}, trust in automation \cite{VanD25}, and explanations in safety-critical decision-making \cite{Murra22,Merwe24,Oestm25}. While technical research on MASS has progressed rapidly \cite{Kim22}, human-centered integration remains an open challenge \cite{Liao20,Liao22,Merwe24a,Kim26,Vasan25}. In particular, existing measures of trust \cite{Mayer95}, attitudes toward automation \cite{Wilso23}, and explanation satisfaction \cite{Hoffm18,Hoffm23} may not fully capture maritime operators’ concerns.

The present study contributes to human-centered AI in the maritime domain by examining how maritime stakeholders perceive trust, technology anxiety, explanation quality, and the perceived advantages and disadvantages of a maritime assistant. The results provide insight into the challenges of measuring trust and explanation needs in a safety-critical domain and highlight the importance of domain-sensitive evaluation methods for future MASS decision-support systems \cite{Longo24,Seebe20}.
Our evaluation scripts and resultant images are available at \url{https://github.com/djirak/AHOI_Survey_HumanMachineTeaming.git}.

\section{Explainable AI in Collision-Avoidance at Sea} 
While MASS and maritime autonomy is progressing, it is inevitably necessary to study needs and demands from the actual seafarers, enabling genuine human-centered AI. Technical success does not measure user comprehension, human cognitive (work)load, and trust calibration, yet they are important when transferring digital solutions into real maritime operational life. To capture the necessary domain knowledge, we conducted a survey study addressed at seafarers with command vessel bridge experiences, such as Captains or Officer-of-the-Watch (OOW). We prepared two collision-avoidance tasks, the first one serving as a ``warm up'', i.e. an unambiguous vessel encounter situation with clear rule following, to ensure that the participants understand the task and the role of the AI. The second scenario represented an ambiguous traffic situation (see section \ref{ssec:survey_flow}). The survey design was driven by the following research questions:
\begin{itemize}
    \item RQ1: What is the seafarer's disposition to technological progress and their propensity to trust? (ATAS scale, pre-questionnaire)
    \item RQ2: To what extend does the inclusion of explainability affect user satisfaction, trust, and willingness to adopt such systems? (TiA, altered Hoffman, post-questionnaire)
    \item RQ3: How do maritime professionals perceive the usefulness and trustworthiness of the maritime assistant? What are the concerns and expectations of seafarers regarding the integration of AI into maritime daily routines and team workflows? (open-question format)
\end{itemize}
Before dissemination, we pilot-tested the the survey among ten professional seafarers asking for their feedback on flow and response time, which resulted in some modifications of images sizes and correct maritime wording. Specifically, we decided to term the XAI as ``maritime assistant'' to avoid confusion and possible negative affect. We disseminated the online study using Qualtrics among professional and nautical students from the Antwerp Maritime Academy (AMA) and using their alumni network. 

\subsection{Survey Flow and Tasks}
\label{ssec:survey_flow}
\begin{figure}[htbp]
\centering
\includegraphics[width=0.8\textwidth]{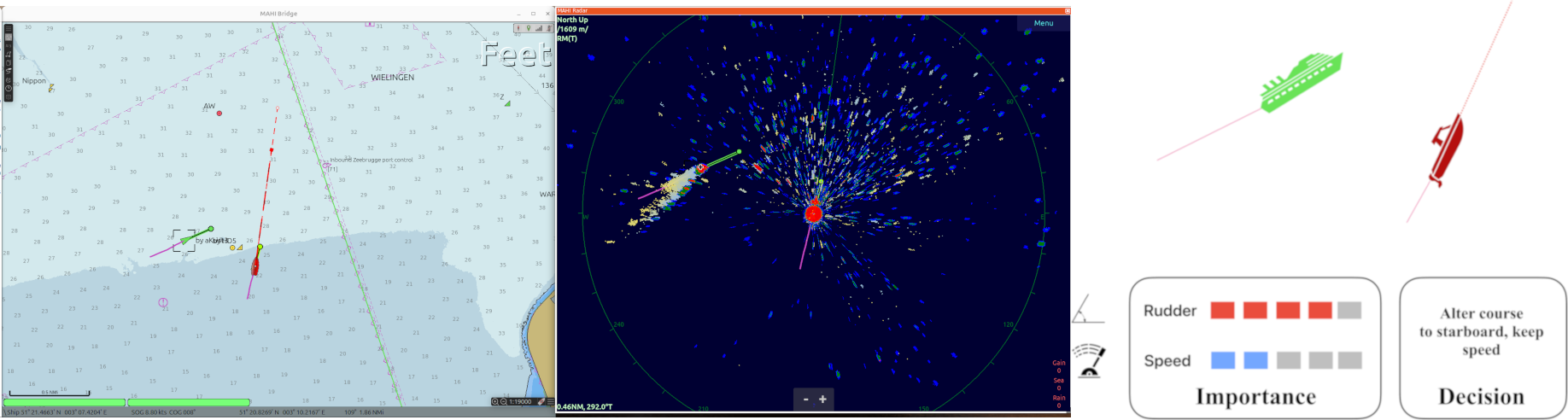}
\caption{Example from Scenario 1: The first two images (ECDIS; AIS) were initially shown without XAI to obtain participants’ independent assessment and decision. They were then presented again with a simplified XAI depiction showing the key decision elements (rudder/speed features) and the agent’s decision} 
\label{fig:scenario1_1}
\end{figure}

\begin{figure}[htbp]
\centering
\includegraphics[width=0.8\textwidth]{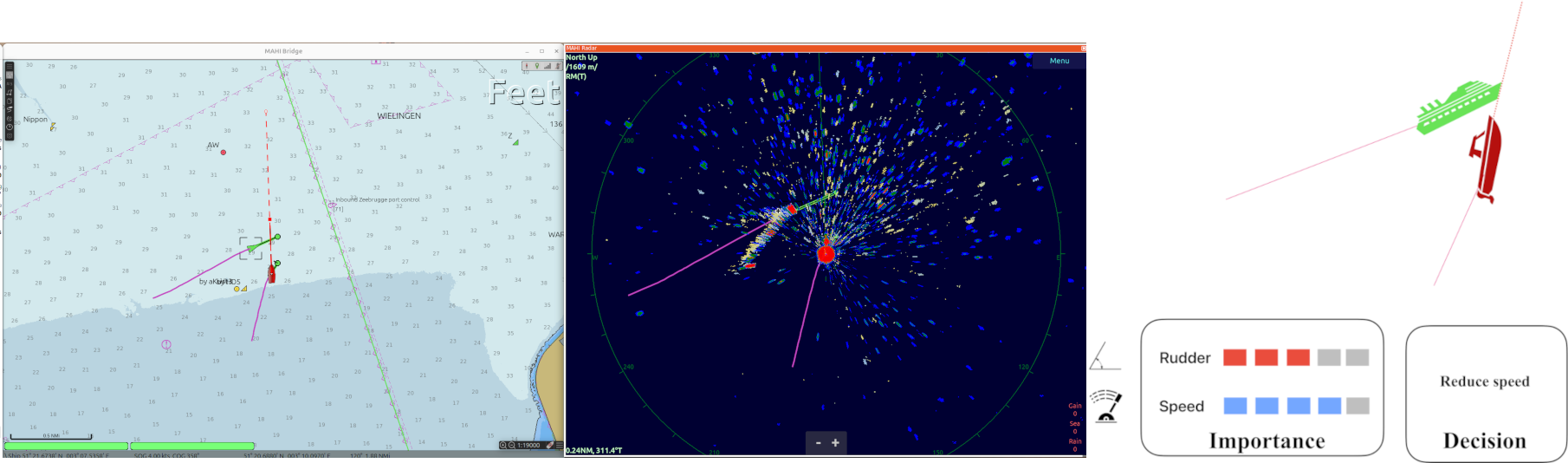}
\caption{Example image from scenario 1, here with a close encounter and different XAI.} 
\label{fig:scenario1_3}
\end{figure}

\begin{figure}[htbp]
\centering
\includegraphics[width=0.8\textwidth]{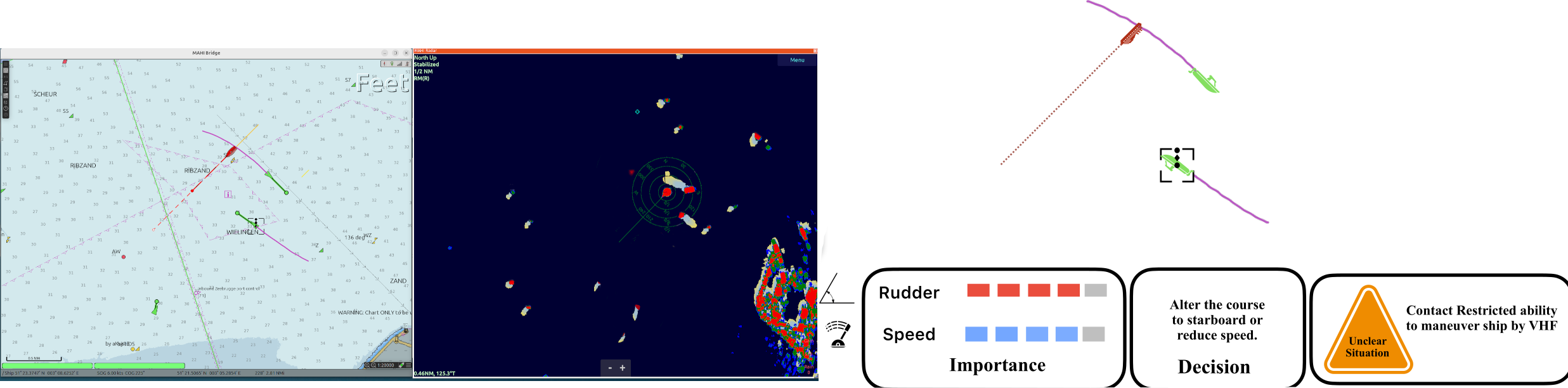}
\caption{Example image from scenario 2. The situation is ambiguous as four other vessels (green), a fisher boat, and a buoy are in close vicinity.} 
\label{fig:scenario2_1}
\end{figure}

\begin{figure}[htbp]
\centering
\includegraphics[width=0.8\textwidth]{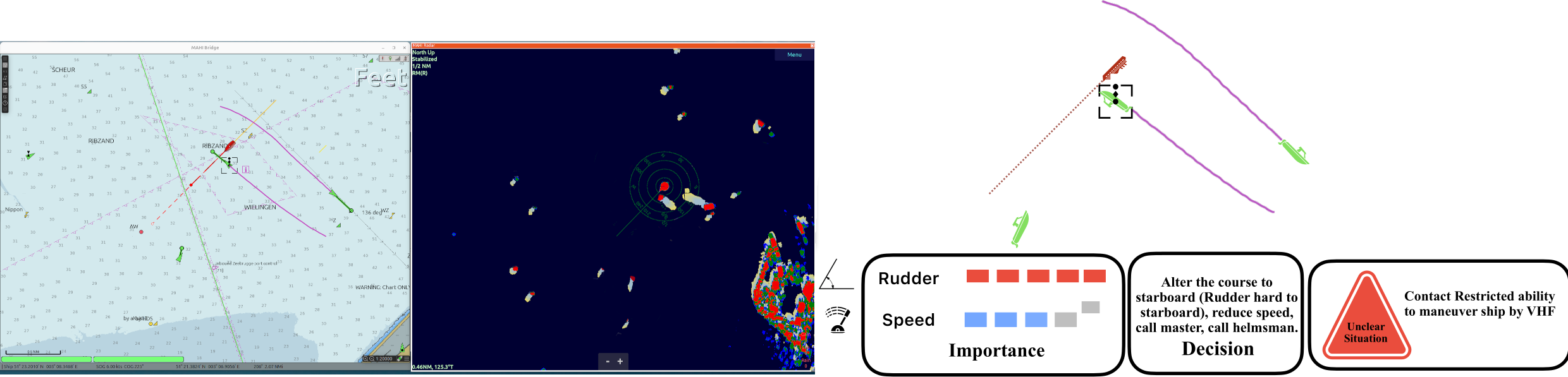}
\caption{Example image from scenario 2 with an additional urgency by XAI to take action.} 
\label{fig:scenario2_4}
\end{figure}

The survey was structured into a briefing phase where participants were informed about the study and gave their consent, followed by a pre-\\questionnaire, the ``Abbreviated Technology Anxiety Scale'' (ATAS \cite{Wilso23}). Afterwards, we presented two collision-avoidance scenarios comprising ECDIS (Electronic Chart Display and Information Systems) and AIS (Automatic Identification System) images created by the Belgian MAHI company \footnote{\url{https://mahi.be/}}. For the XAI, we created images aligning the actual situation with the correct decision. We did not introduce deceptive agent behavior. 
Afterwards, we started with the presentation of two collision-avoidance tasks, following a ``Wizard-of-Oz'' approach (WoZ), i.e. simulating collision encounters. Figures \ref{fig:scenario1_1}-\ref{fig:scenario2_4} show example images as originally shown in the survey. The first scenario represented regular encounter at sea with clear procedure. We first presented the ECDIS and AIS images (first two images in Figures \ref{fig:scenario1_1}-\ref{fig:scenario1_3}) to the participants and asked them to assess the situation: \textit{``Assume you are the Master of the red vessel. The OOW delayed calling you for an unclear close quarter situation. When you arrive at the bridge, you see the present situation''}. The response type had been provided in an open format (\textit{``Please describe what you notice, what concerns you, an what action you would consider at this moment.''}). In total, we showed four traffic situations simulating a close encounter. Afterwards, we presented the same scenario but with an additional XAI output displaying the vessels' parameters 'speed' and 'rudder' with respective importance scores influencing the decision. After the completion of scenario 1, we asked the participants to fill out the ``Trust in Automation'' scale (TiA \cite{Jian00}) and a modified version of the Hoffman XAI scale (see Figure \ref{fig:hoffman}) \cite{Hoffm18}. We then proceeded with the presentation of four images for scenario 2, exemplary shown Figures \ref{fig:scenario2_1}-\ref{fig:scenario2_4}, which represented dense traffic with four additional vessels (in green), a fisher boat, and a buoy in close vicinity, thus a more ambiguous situation than for scenario 1. Both scenarios and the correct interpretation of situations or ``ground truth'' (e.g., COLREG rules) were provided by Cpt. Armeen Saroukanoff. Afterwards, the participant filled out again the TiA and altered Hoffman scale.

After the scenarios, participants provided open-ended feedback on the perceived advantages and disadvantages of an autonomous maritime agent. They then reported demographic information, including age, sea experience, rank (OOW, Captain, pilot, student), and education. Age and sea experience were interval-coded into five groups.

\subsection{Survey Collection and Demographics}
Anonymous responses were collected from December 2025 to mid-April 2026, with duplicate checks to avoid redundancy. In total, 166 persons accessed the survey; after filtering for completed responses, 66 remained. Metadata indicated that most dropouts occurred within the first ten minutes, suggesting that survey fatigue was unlikely the main reason. The filtered sample showed an uneven demographic distribution, as illustrated in Table \ref{tab:demographics}, with two dominant age cohorts (groups 2 and 5) and about half of the respondents reporting at least 72 months of sea service. Analysis on the `ranks' further showed that approximately half of the participants were Captains or Officers of the Watch (OOWs), indicating a high level of maritime expertise captured in the survey.

For the completed responses, participants’ observations and decisions were compared with the expected scenario interpretations. Overall, responses aligned with the intended observations and behaviors, which was expected since the scenarios did not include unrealistic or deceptive elements.
\begin{table}
    \centering
        \caption{Demographics of the survey respondents. Note that the total numbers for 'age' are 66, while only 62 participants provided their sea experience.}
    \begin{tabular}{ccccc}\toprule
         Group&  Age & \#Respondents&  Sea Experience (months)& \#Respondents\\\midrule
         1&   18-24&10&  0-3& 7\\
         2&   25-34&17&  4-12& 5\\
         3&   35-44&10&  13-36& 9\\
         4&   45-54&10&  37-72& 10\\
         5&   >54&19&  >72& 31\\ \bottomrule
    \end{tabular}
    \label{tab:demographics}
\end{table}

\section{Results}
We first analyze RQ1\&RQ2 from the questionnaires for a quantitative evaluation. We then synthesize the results from the open format feedback by sentiment analysis and thematic clustering (RQ3) in section \ref{ssec:open_format}. 

\subsection{Questionnaires Evaluation}
To address RQ1, we used questionnaires to quantify participants’ technology disposition and trust. As a pre-questionnaire, we administered the 11-item Abbreviated Technology Anxiety Scale (ATAS) \cite{Wilso23}, which captures technology anxiety through negatively worded statements such as \textit{``I am not a technology person''}. The scale showed good internal consistency (Cronbach’s $\alpha = 0.86$).

\begin{figure}[htbp]
\centering
\includegraphics[width=0.8\textwidth]{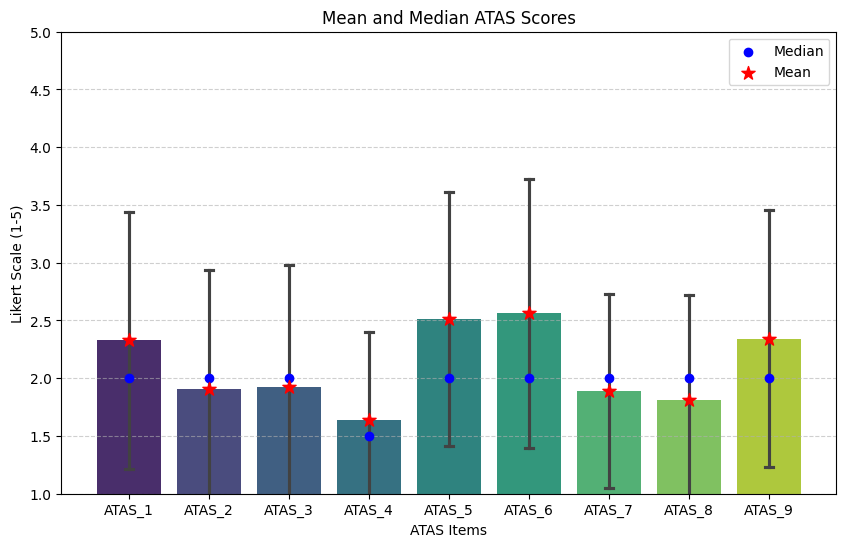}
\caption{Score distribution for the Abbreviated Technology Anxiety Scale (ATAS \cite{Wilso23}).} 
\label{fig:atas}
\end{figure}

Participants rated each item on a 5-point Likert scale (1 = ``Strongly Disagree'', 5 = ``Strongly Agree''), with higher scores indicating greater technology anxiety. As shown in Figure \ref{fig:atas}, the predominant median score was 2, suggesting that most participants disagreed with the ATAS statements and generally held positive attitudes toward technology. However, items 5 (\textit{`I feel technology complicates simple tasks''}) and 6 (\textit{`Keeping up with the newest technology is impossible''}) showed discrepancies between mean and median scores, as well as higher standard deviations. This may indicate concern about future technology integration despite general openness toward new systems, which is plausible given maritime personnel’s regular exposure to new bridge instruments and software. Similarly, item 9 (\textit{``I am often annoyed when using technology''}) showed higher scores and variance, suggesting that some participants associate technology use with frustration, usability challenges, or the effort required to adapt to new systems.

After each collision-avoidance scenario, participants completed the Trust in Automation (TiA) questionnaire \cite{Jian00}. The original 12-item scale was reduced to 11 items by omitting the familiarity item. Items 6-11 were reverse-coded so that higher scores consistently indicated greater trust in automation. For consistency with the other survey instruments, all TiA items were administered on a 5-point Likert scale. The scale showed high reliability in both scenarios (scenario 1: $\alpha=0.910$; scenario 2: $\alpha=0.936$).

\begin{figure}[htbp]
\centering
\includegraphics[width=0.9\textwidth]{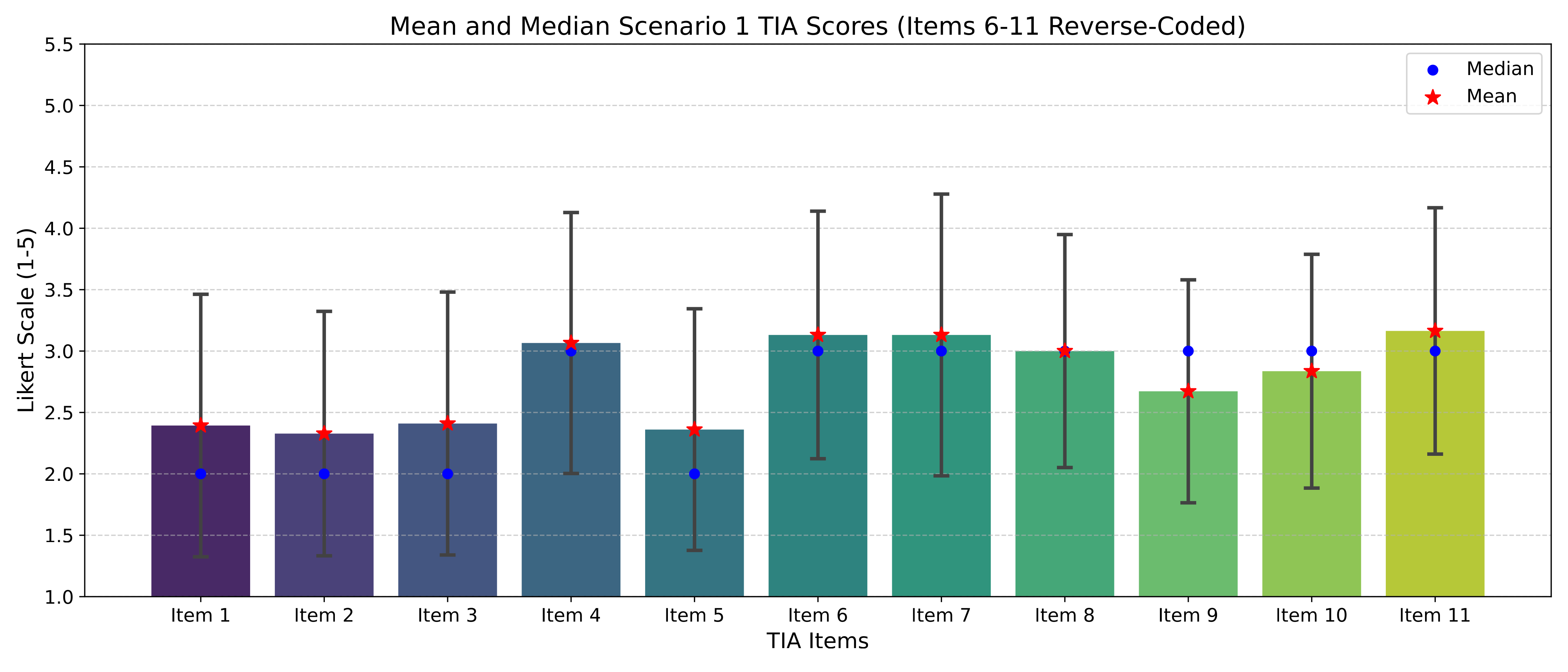}
\caption{Results from Trust in Automation Scale (TiA) for Scenario 1.} 
\label{fig:tia1}
\end{figure}

\begin{figure}[htbp]
\centering
\includegraphics[width=0.9\textwidth]{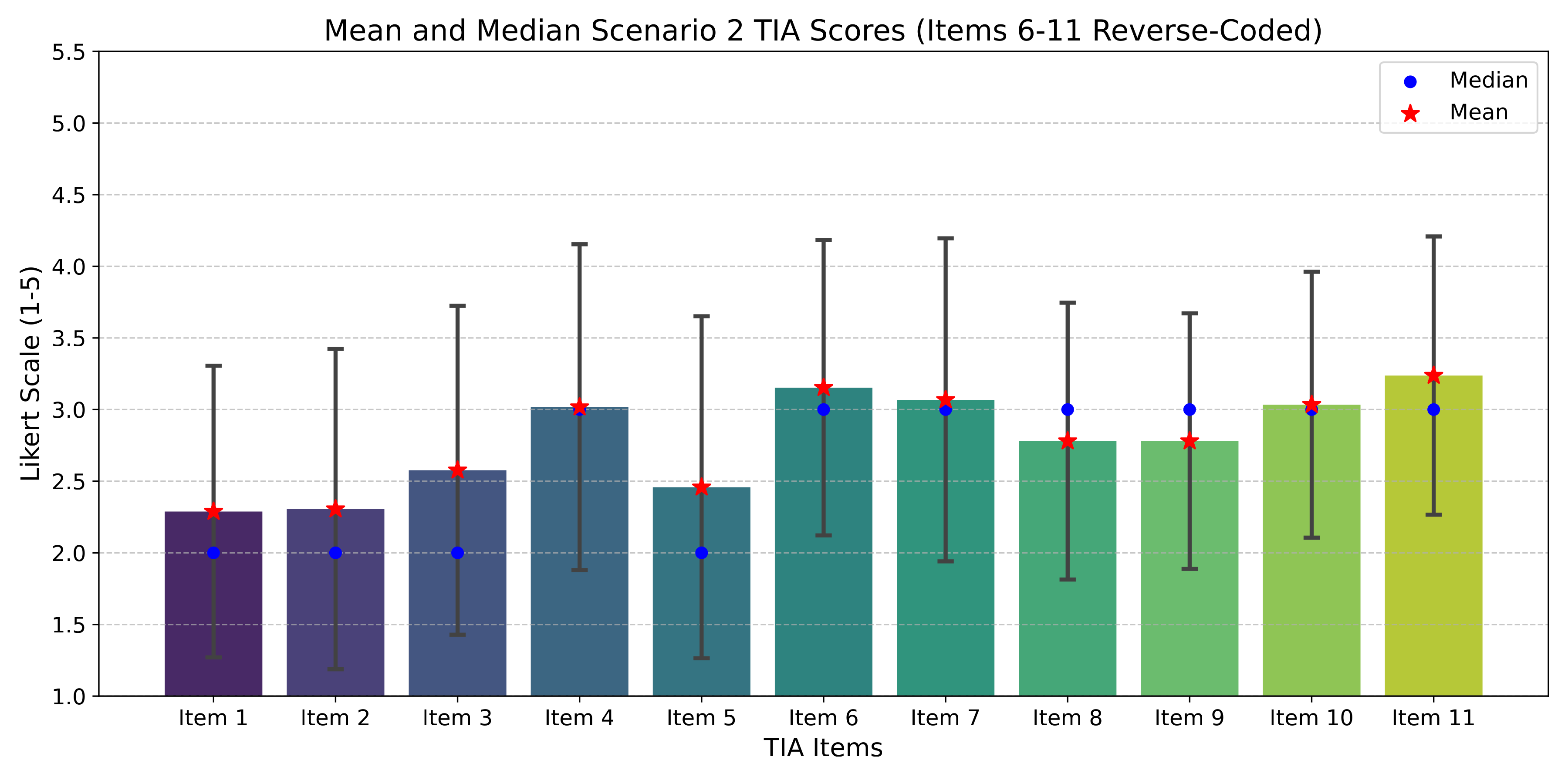}
\caption{Results from Trust in Automation Scale (TiA) for Scenario 2.} 
\label{fig:tia2}
\end{figure}

Figures \ref{fig:tia1}-\ref{fig:tia2} show similar TiA response patterns across both scenarios. A paired-samples t-test indicated no significant difference between scenario 1 and scenario 2 ($p = 0.936 > 0.05$), suggesting that overall trust in automation remained stable across scenarios.

Overall, participants showed a moderately favorable but cautious perception of the assistant. For the negatively worded items (1-5), the median score of 2 suggests that participants generally disagreed with adverse characterizations of the assistant, such as being unreliable, deceptive, or suspicious, although these concerns were not entirely absent. For the positively worded items (6-11), the median score of 3 indicates a more neutral opinion, i.e. participants did not reject positive trust-related statements, but also did not strongly endorse them. This suggests that participants were not distrustful of the assistant, but that their trust was not yet strongly consolidated and may depend on further experience, demonstrated reliability, and integration into bridge routines.

We tested whether older or more experienced participants showed higher ATAS scores. Since both variables were interval-coded, Kruskal-Wallis tests were used to compare ATAS responses across groups. No significant differences were found for age ($H=2.150$, $p=0.708$) or sea experience ($H=5.747$, $p=0.219$) at $\alpha=0.05$. This suggests that technology anxiety did not differ significantly across age or experience groups, supporting the overall finding that respondents were generally open toward technology.

\begin{figure}[htbp]
\centering
\includegraphics[width=0.9\textwidth]{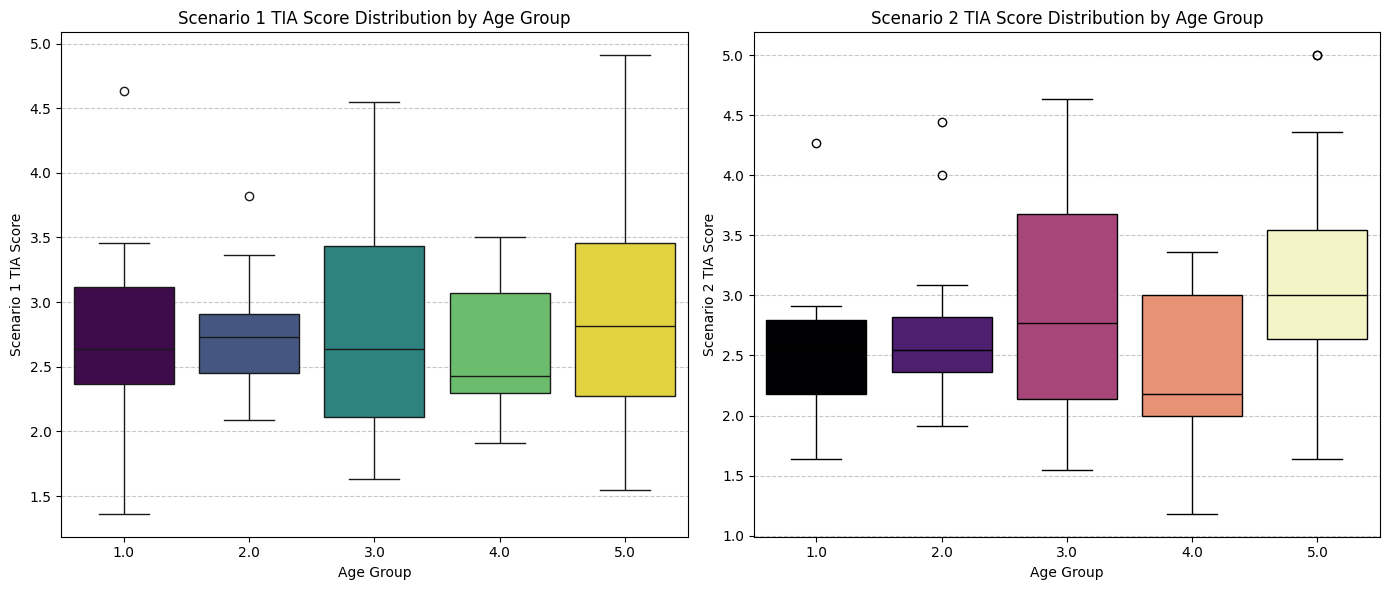}
\caption{Distribution of TiA scores vs. age groups. Both descriptive and the Kruskal-Wallis test do not confirm a possible correlation between different age groups and trust in automation.} 
\label{fig:tia_vs_age}
\end{figure}

Kruskal-Wallis tests showed no statistically significant differences in TiA scores across sea-experience groups for either scenario 1 ($H=4.197$, $p=0.380$) or scenario 2 ($H=5.786$, $p=0.216$). Similarly, no significant age-group differences were found for scenario 1 ($H=0.606$, $p=0.962$) or scenario 2 ($H=8.969$, $p=0.062$). Thus, the results do not support the hypothesis that older or more experienced maritime professionals show lower trust in automation. However, the higher $H$ statistic and near-significant $p$ value in scenario 2 suggest more pronounced age-group differences under increased traffic complexity, as also reflected in Figure \ref{fig:tia_vs_age}. Higher within-group variance, particularly in age groups 3 and 5, further indicates heterogeneous responses. Overall, the present data do not allow statistically robust conclusions about the influence of age or sea experience on trust in automation.

\begin{figure}[htbp]
\centering
\includegraphics[width=0.8\textwidth]{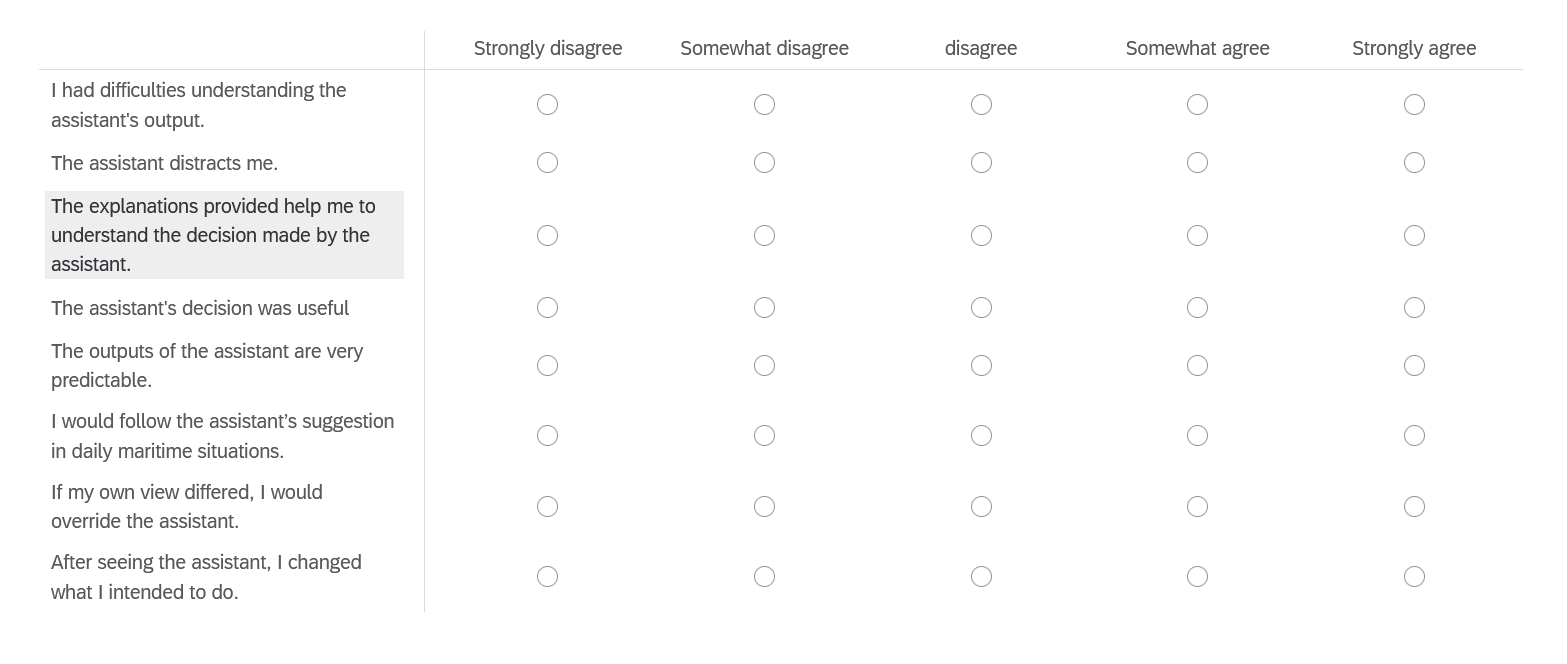}
\caption{Altered Hoffman scale items for explainability effects on perception and decision-making after exposition to the maritime assistant.} 
\label{fig:hoffman}
\end{figure}

\begin{figure}[htbp]
\centering
\includegraphics[width=0.9\textwidth]{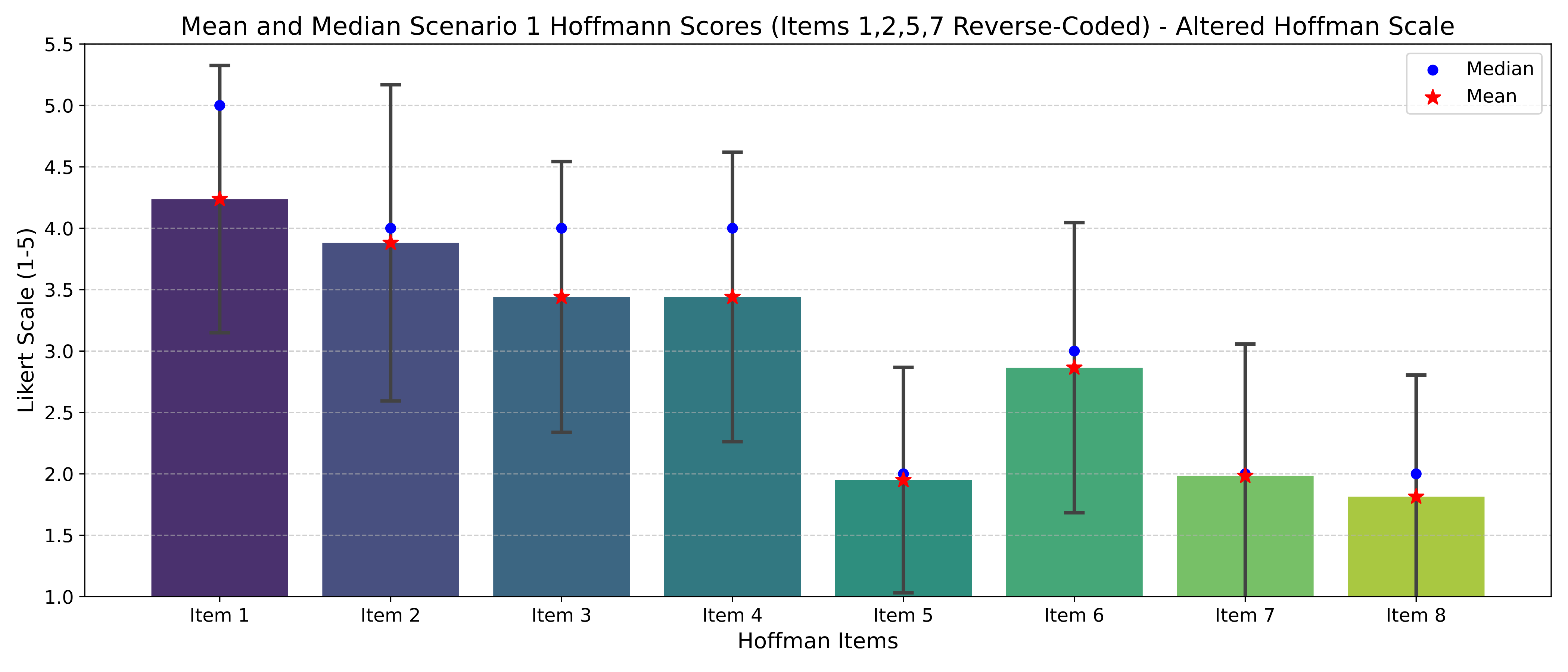}
\caption{Response distribution of the altered Hoffman scale for scenario 1.} 
\label{fig:hoff_scenario1}
\end{figure}

\begin{figure}[htbp]
\centering
\includegraphics[width=0.9\textwidth]{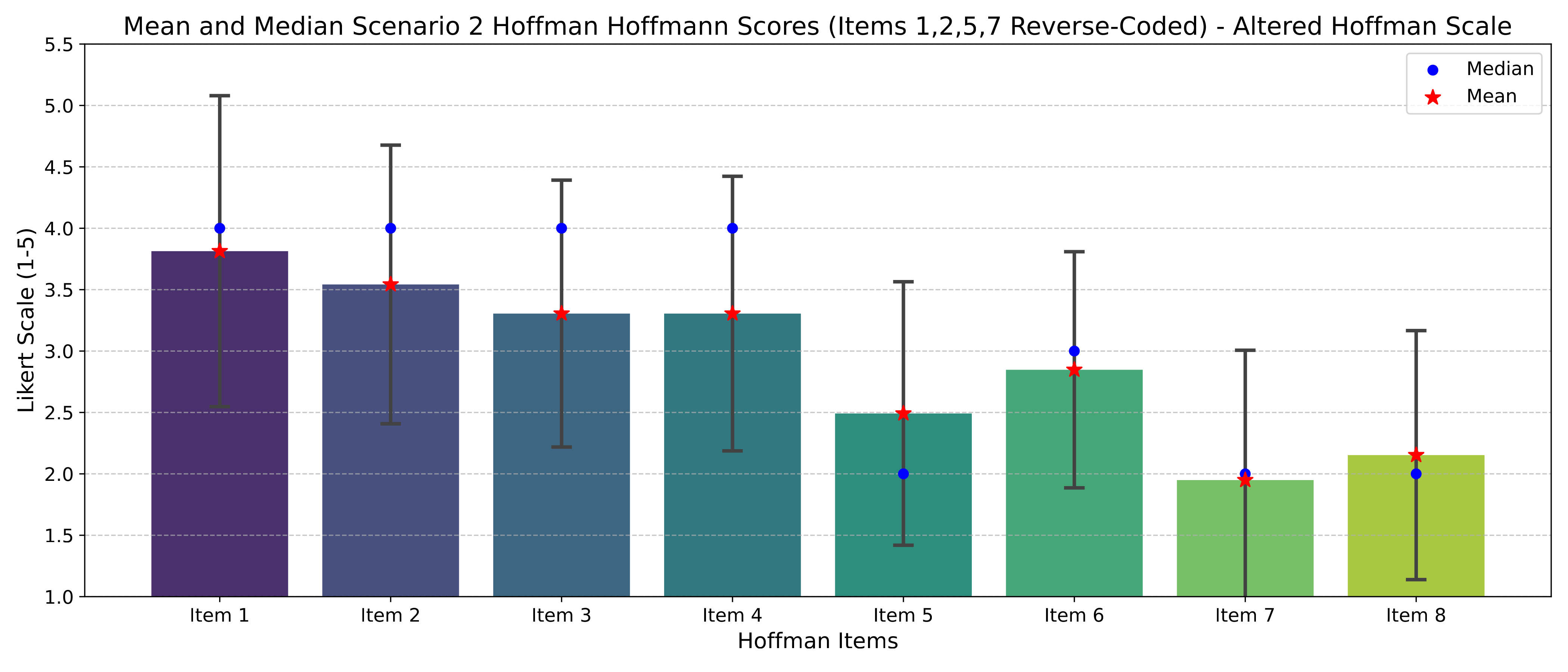}
\caption{Response distribution of the altered Hoffman scale for scenario 2.} 
\label{fig:hoff_scenario2}
\end{figure}

To answer RQ2, we extracted items from the XAI scale provided by Hoffman et al. \cite{Hoffm18,Hoffm23}, shown in Figure \ref{fig:hoffman}. We reverse-coded items 1,2,5, and 7 based on the wording, i.e. item 1 and 2 are negatively worded, and a correlation analysis showing negative correlation of item 5 and 7 with respect to the other items. After this step,  we computed Cronbach's $\alpha=0.633$ for scenario 1 and $\alpha=0.609$ for scenario 2, which represents a moderate internal scale consistency. To understand the divergence of Cronbach's $\alpha$ from ATAS and TiA, we subsequently performed a factor analysis which  revealed three distinct latent factors (eigenvalue >1). This means hat the scale measures a multi-dimensional construct rather than one specific motif. However, we want to provide the descriptive statistic similar to ATAS and TiA to complete the picture and providing at least a trend exhibited by the respondents. Figure \ref{fig:hoff_scenario1} and Figure \ref{fig:hoff_scenario2} show the response distribution based on a 5-point Likert scale. Due to the reverse-coding, items 1,2, 5, and 7 are positively interpreted. For example item 1: \textit{``I had difficulties understanding the assistant's output''} with an original mean of $1.7$, i.e. disagreement on this statement, has now an average response score of $4.1$, meaning agreement that the assistant's output was easy to understand. 

Overall, only minor response differences were observed between scenarios 1\&2, suggesting that increased traffic complexity did not negatively affect the perceived agency or reliability of the agent. This was supported by a paired t-test, which showed no statistically significant difference. Interestingly, participants were unlikely to change their decision based on the agent’s output, as indicated by a median score of 2 for item 8. A slight shift toward agreement in Scenario 2 may suggest that seafarers are more inclined to consult the assistant in ambiguous traffic situations. Items 3, 4, and 5 showed the highest variance, possibly reflecting differing perceptions of the explanation design and usefulness, as well as limitations of the Wizard-of-Oz setup, which may have lacked the realism needed for robust system judgments.

\subsection{Qualitative Analysis and Synthesis of Open Format Responses}
\label{ssec:open_format}
Responses from the participant in the open format were analyzed using a two-step procedure. First, sentiment scores were computed to obtain an overview of positively and negatively connoted responses. Second, responses were manually inspected and grouped into recurring themes reflecting perceived advantages and disadvantages of the maritime assistant. Sentiment scores were used to guide the identification of strongly positive and negative responses, but themes were derived from qualitative interpretation of the response content. Direct quotations were selected based on thematic clarity and sentiment polarity. Table \ref{tab:themes} displays themes we identified and grouped regarding perceived advantages and disadvantages of the maritime assistant. 
\begin{table}[h]
    \centering
    \caption{Summary of major themes for (dis)advantages of the ``maritime assistant''}
    \label{tab:themes}
    \begin{tabular}{p{2cm}p{3cm}p{6cm}}
        \toprule
        \textbf{Sentiment} & \textbf{Theme}  & \textbf{Quote} \\
        \midrule
        Advantage & Confidence support& ``\dots for young officers of the watch this will give them confidence in making the safest decision.'' \\
        Advantage & Situation awareness  & ``It is just an extra tool to help the OOW build a good situation awareness.'' \\
        Advantage & Time-saving &  ``The time saved can be used to perform a deeper analysis taking into account external factors that the assistant could not necessarily have\dots'' \\
        Disadvantage & Over-reliance & ``\dots the OOW would stop thinking by himself\dots'' ``\dots overdependence and loss of skill is a risk in my opinion and might lead to lowering of standards\dots'' \\
        Disadvantage & Skill degradation  & ``\dots Junior officers no longer learn to judge traffic situations by their own \dots'' \\
        Disadvantage & Input reliability & ``\dots if sensors fail, or the input is faulty \dots the decisions of the assistant can be wrong''\\
        \bottomrule
    \end{tabular}
\end{table}
Responses on disadvantages revealed several safety-critical concerns. Participants highlighted risks of erroneous recommendations due to faulty sensor input, GPS spoofing, jamming, or other input failures. They also questioned whether the assistant could handle complex, sequential traffic situations rather than isolated encounters. Beyond technical reliability, a central concern was human over-reliance, particularly among junior OOWs, which could reduce vigilance, traffic judgment, and long-term skill development. Additional concerns included information overload, distraction, and alarm burden on the bridge. Thus, negative sentiment did not indicate a general rejection of AI support, but rather concern about the conditions under which such support could become unsafe.

We further analyzed feedback comments through sentiment analysis, interpreting positive comments as perceived opportunities and negative comments as perceived risks. As with the themes in Table \ref{tab:themes}, comments were grouped into higher-level themes and synthesized in Table \ref{tab:synthesis}. Although some themes appear contradictory, such as improved versus reduced situation awareness, they highlight central design trade-offs for XAI and broader UI/UX development. In this context, the key challenge is not only AI performance, but the design of user-centred interfaces that avoid distraction, support natural communication, and fit maritime routines. Overall, the comments show that maritime personnel are aware of both opportunities and risks, supporting the quantitative finding of general openness toward new technology while providing domain-specific insights that are highly valuable for researchers developing for but not necessarily with experts in the maritime domain.

\begin{table}[h]
    \centering
    \caption{Synthesis of chances and risks provided by the open feedback question}
    \label{tab:synthesis}
    \begin{tabular}{p{3cm} p{4.5cm} p{4.5cm}}
        \toprule
        \textbf{Function} & \textbf{Opportunity} & \textbf{Risk} \\
        \midrule
        Decision recommendation & Supports faster and potentially safer decision-making & May encourage blind following or wrong maneuvers \\
        Confidence support & Helps young or uncertain officers validate decisions & May lead to overconfidence or dependency \\
        Situation awareness support & Helps organize traffic information & May add distraction or information overload \\
        Rule-based reasoning & COLREG alignment increases credibility & Wrong input may still produce wrong rule-based advice \\
        Predictive support & Helps anticipate traffic conflicts in advance & May fail in complex or sequential encounters \\
        Training support & Can support learning and confidence-building & May reduce development of independent judgment \\
        Automation support & Reduces workload in standard situations & May create complacency in non-standard situations \\
        \bottomrule
    \end{tabular}
\end{table}

Sentiment analysis showed age-related differences in how participants described perceived advantages and disadvantages. Across all age groups, comments on advantages were generally positive, indicating that participants used favorable language when discussing beneficial aspects of the assistant. The highest average positive sentiment was observed in age group 4 (45-54 years), while age group 5 ($>54$ years) also remained positive, though slightly less so than age groups 3 and 4.

For disadvantages, the pattern was more varied. Age groups 1, 2, and 4 showed negative average sentiment, suggesting more critical language when discussing drawbacks. In contrast, age groups 3 and 5 showed slightly positive average sentiment even for disadvantages. This was particularly notable for the oldest group, indicating that these participants described disadvantages in comparatively less negative terms than several younger groups.

\section{Discussion}
The development of Maritime Autonomous Surface Ships (MASS) \cite{Li24} and AI-supported navigation marks an important transition in maritime operations, where decision-making is increasingly shared between human operators and intelligent systems \cite{Liao20,DeVos21,Murra22,Merwe26}. However, autonomy in safety-critical domains cannot be evaluated solely as a technical capability. Its successful integration depends on how well AI systems fit the operational realities, responsibilities, and decision practices of maritime professionals \cite{Murra22}. 

This survey provided unique insights into maritime AI and autonomy from the perspective of domain experts and experienced maritime personnel. Overall, participants showed a generally positive attitude toward new maritime technologies and recognized the potential of AI-supported assistants to support decision-making, situation awareness, and operational efficiency (RQ1-RQ2). At the same time, the findings indicate that future development should not aim at increasing trust unconditionally, but at supporting calibrated trust through reliable, transparent \cite{Malla23,Merwe23}, and context-sensitive system behavior and explanations (XAI)\cite{Longo24,Ranja25}. Key challenges therefore lie in the joint design of autonomous agents and XAI interfaces that make system capabilities, limitations, and uncertainties understandable to operators (RQ3). 
Contrary to a common assumption that older operators may be less open to technological innovation, the analysis did not reveal clear age-related differences in technology openness. Participants across age groups showed broadly comparable attitudes, indicating that age alone may not be a sufficient predictor of acceptance or skepticism toward maritime AI assistance. In this context, openness to technology may be more strongly influenced by domain-specific factors, such as perceived usefulness, responsibility for safety-critical decisions, and familiarity with bridge systems. This suggests that future evaluations of maritime AI should avoid treating age as a proxy for technological openness and instead consider the operational and experiential factors that shape human-AI interaction in maritime practice.

Several limitations should be considered when interpreting the results. First, the survey showed inconsistent response patterns and a substantial reduction from accessed to completed responses, resulting in a limited final sample size. This restricts the generalization of the findings and indicates that the survey design should be revised in future studies. In particular, static scenario images may not have provided a sufficiently realistic or engaging representation of maritime decision-making, which may have contributed to early dropouts or reduced immersion. Future surveys could therefore use short scenario videos or lightweight online demonstrations to provide a more dynamic and realistic impression of the assistant and its explanations. Beyond survey refinement, future work should focus on the design and evaluation of maritime AI agents and XAI interfaces together with domain experts. A central challenge will be to identify the appropriate balance between providing useful explanations, avoiding cognitive overload and alarm fatigue, and supporting calibrated trust in autonomous agents.

\section{Conclusion}
This study contributes to the human-centred evaluation of maritime autonomy by showing that maritime stakeholders are generally open to AI-supported decision assistance, while remaining attentive to risks of reliability, over-reliance, distraction, and loss of expertise. The findings suggest that future MASS and maritime AI systems should not focus solely on increasing automation or trust, but on supporting calibrated reliance through transparent, reliable, and operationally meaningful design. Human expertise should therefore remain central to the development and evaluation of maritime autonomous systems.

\begin{credits}
\subsubsection{\ackname} The authors thank the MAHI company for the provision of the radar images and Arian Sabaghi Khameneh (imec/IDLab) for the provision of the XAI images. This work has been conducted within the DEFRA AHOI project, funded by the Belgian Royal Higher Institute for Defence, under contract number 23DEFRA002.

\subsubsection{\discintname}
The authors have no competing interests to declare that are
relevant to the content of this article.
\end{credits}
%
%
%
%
\bibliographystyle{splncs04}
\bibliography{hldm26}

\end{document}